# Excitation and detection of coherent sub-terahertz magnons in ferromagnetic and antiferromagnetic heterostructures

Shihao Zhuang and Jia-Mian Hu[*]

*Department of Materials Science and Engineering, University of Wisconsin-Madison, Madison, WI, 53706, USA*

**Abstract**

Excitation of coherent high-frequency magnons (quanta of spin waves) is critical to the development of high-speed magnonic devices. Here we computationally demonstrate the excitation of coherent sub-terahertz (THz) magnons in ferromagnetic (FM) and antiferromagnetic (AFM) thin films by a photoinduced picosecond acoustic pulse. Analytical calculations are also performed to reveal the magnon excitation mechanism. Through spin pumping and spin-charge conversion, these magnons can inject sub-THz charge current into an adjacent heavy-metal film which in turn emits electromagnetic (EM) waves. Using a dynamical phase-field model that considers the coupled dynamics of acoustic waves, spin waves, and EM waves, we show that the emitted EM wave retains the spectral information of all the sub-THz magnon modes and has a sufficiently large amplitude for near-field detection. These predictions indicate that the excitation and detection of sub-THz magnons can be realized in rationally designed FM or AFM thin-film heterostructures via ultrafast optical-pump THz-emission-probe spectroscopy.

[*]E-mail: jhu238@wisc.edu

**Introduction**

Ultrafast magnetoacoustics is focused on studying the interaction of femtosecond (*fs*)-laser-induced ultrashort (picosecond) acoustic pulse with magnetic order[1]. It allows for investigating the magnetoelastic coupling at a possibly fastest attainable timescale and offers opportunities for the design of high-speed, compact, and energy-efficient magnetic devices. Due to the time compression of the laser energy, such photoinduced picosecond (*ps*) acoustic pulse can have an ultrahigh strain amplitude of >1%[2] and short wavelengths down to a few nanometers (nm). In principle, such large and non-uniform strain would excite short-wavelength (high-frequency) magnons (quanta of spin waves) via magnetoelastic coupling[3,4], which underpin the development of magnonic logic circuits[5,6] operating at sub-terahertz (THz) frequencies that are much faster than existing gigahertz (GHz) circuits. However, existing experiments in ultrafast magnetoacoustics are mostly limited to the observation of small-angle precession of averaged magnetization in ferromagnets (e.g., (Ga,Mn)As and FeGa) or ferrimagnets (yttrium iron garnet)[7–13]. In these experiments, ultrafast time-resolved magneto-optical Kerr microscopy (TR-MOKE) is utilized to optically pump the optical-to-acoustic transducer (typically a thin-film metal) and detect the temporal change in the average magnetization within the penetration depth (typically a few to tens of nm) of the probe laser pulse. As a result, signals of sub-THz magnons, which also have nm-scale wavelength, could be averaged out.

Theoretical and computational efforts have also been devoted to understanding and predicting ultrafast magnetoacoustic phenomena in ferromagnetic (FM) thin films. These include acoustically induced of precession of uniform magnetization[14–16], and the acoustic excitation of exchange-dominated magnon modes in FM films[17–19]. Specifically, it has been predicted that a single *ps* acoustic pulse can excite standing sub-THz magnon modes in thin (at most tens of nm) FM films[17,18] and propagating THz magnon mode in thick (at least hundreds of nm) FM films[19] along the film thickness direction.

In this paper, we demonstrate, using analytical calculation and dynamical phase-field simulations, the excitation of sub-THz standing magnon modes (frequency: 0.1-1 THz) in both FM and antiferromagnetic (AFM) thin films by a single *ps* acoustic pulse. We also calculate the injection of spin current from the excited sub-THz magnons into an adjacent heavy metal (HM) thin film, the spin-charge current conversion in the HM via the inverse spin Hall effect (iSHE), and the emission of electromagnetic (EM) wave that arises mainly from the charge current. It is found that the emitted EM wave retains the spectral information of all the excited sub-THz magnons, providing a basis for magnon detection by THz emission spectroscopy.

**Results**

*Excitation and detection of sub-THz magnons in ferromagnetic thin film*

We propose a metal/dielectric/FM/HM thin-film heterostructure, as shown in Fig. 1a, using Al/MgAl$_2$O$_4$(001)/MgAl$_{0.5}$Fe$_{1.5}$O$_4$(001)/Pt as an example. Recent computations[17,18] predict that a *ps* acoustic pulse can excite multiple coherent standing magnon modes (*n*=0, 1, 2,…∞) in a FM thin film via magnetoelastic coupling, and that the frequencies of high-order (*n*≥1) magnon modes can reach sub-THz range. Polycrystalline Ni[17] and (001) Fe$_{81}$Ga$_{19}$[18] thin films were utilized as examples. Here, we computationally demonstrate that such sub-THz magnon modes can be detected by adding a HM thin film on top of the FM thin film to enable the spin pumping at the FM/HM interface and spin-charge conversion in the HM layer via the iSHE. As will be shown later, the frequency spectra of both the oscillating charge current in the HM layer and the near-field EM wave emission are largely the same as the magnon spectrum. Therefore, the magnon modes detection can be achieved by measuring the EM wave emission. Since the *ps* acoustic pulse can be obtained by irradiating the metal transducer with a *fs* laser pulse, we suggest that the excitation and detection of sub-THz magnons can be achieved by ultrafast optical-pump THz-emission-probe spectroscopy[20,21], which has successfully been utilized to excite and detect fs-laser-induced THz spin current (incoherent magnons)[22–26] and ultrafast demagnetization[27,28]. It is worth noting that the addition of a spin-charge conversion layer is necessary for the detection of high-order magnon modes ($n \geq 1$), because the net EM wave emission produced directly by the magnons via magnetic dipole radiation is negligible.

We now discuss the principles of materials selection for each layer of the heterostructure. First, polycrystalline Al film is commonly used as an optical-to-acoustic transducer[7–12] due to its large thermal expansion and small absorption length to near-infrared laser. Our previous computation[19] predicts that tuning the Al film thickness allows for effective tuning of both the peak strain amplitude and duration of the acoustic pulse. For example, a bipolar Gaussian-like acoustic pulse with a peak strain amplitude $\varepsilon_{max} \sim 0.3\%$ and a duration of ~10 ps can be obtained in a Al(10nm)/MgO(001) bilayer under the excitation of a single-shot *fs* laser pulse (absorbed fluence: 1.3 mJ/cm$^2$, duration: 20 fs, wavelength: 800 nm)[19]. Here, we directly consider the injection of a bipolar Gaussian acoustic pulse (see Fig. 1a, and the Methods section) into the dielectric layer for simplicity. Second, the dielectric layer can shield the magnetic layer from rapid laser heating and confine the laser-excited hot electrons within the metal transducer to obtain acoustic pulse of large strain amplitude. Third, the magnetic layer needs to simultaneously have large magnetoelastic coupling and low magnetic damping, which lead to large amplitude and long duration of the excited magnons. Moreover, the magnetic layer is also best to be an electronic insulator to suppress the eddy current loss. Furthermore, the magnetic layer needs to sufficiently thin to excite high-frequency magnon modes, as will be elaborated later. Considering all these aspects, a promising material candidate is the spinel ferrite film (001)MgAl$_{0.5}$Fe$_{1.5}$O$_4$ (MAFO), which can be epitaxially grown on (001)MgAl$_2$O$_4$ (MAO) substrate. As a magnetic insulator, structurally coherent MAFO thin films[29] (thickness < 20 nm) simultaneously have a large magnetoelastic coupling coefficient $B_1 \sim 1.2$ MJ m$^{-3}$ (c.f., $B_1 \sim 0.3$ MJ m$^{-3}$ in yttrium iron garnet, or YIG), a low Gilbert magnetic damping of $\alpha \sim 0.0015$ (c.f., $\alpha \sim 10^{-4}$ for YIG). By comparison, the large unit cell of YIG (1.24 nm) makes it challenging to grow high-quality thin film[30]. Lastly, the HM layer needs to have large spin-mixing conductance at the magnet/HM interface and large spin Hall ratio to achieve efficient spin-to-charge current conversion. In this regard, crystalline Pt film, which is shown to display a large spin Hall ratio[31] of ~0.8 when grown on MAFO film, is considered. In this work, the Pt thickness is set to be 7 nm, about two times larger than its spin diffusion length[31,32], which is thick enough to absorb almost all the injected spin current and thin enough to maintain a low eddy current loss.

An analytical theory on acoustic excitation of exchange-dominated standing magnons in isotropic FM thin films was presented in ref. [17], where the magnetocrystalline anisotropy was dropped. Here we derive an analytical expression of the magnon dispersion relation $\omega(k)$ in FM thin films with cubic magnetocrystalline anisotropy (see Methods), where $k$ is the angular wavevector along the film thickness direction and $\omega$ is the angular frequency of the magnon. For standing magnons, $k=n\pi/d$ ($n$=0, 1, 2, … ∞) where $d$ is the FM film thickness. This allows us to analytically calculate the frequency $f$ ($=\omega/2\pi$) of each magnon mode as a function of $d$. Note that frequency of $n$=0 mode magnon is also the ferromagnetic resonance (FMR) frequency. The calculation results for (001)MAFO films are shown in Fig. 1b, which provide guidance on the acoustically mediated magnon excitation. For example, let us assume the MAFO film thickness is 15 nm and that the frequency window (illustrated as the vertical dashed line in Fig. 1b) contains up to 500 GHz. In this case, three magnon modes, with frequency of 422.8 GHz ($n$=2), 106.5 GHz ($n$=1), and 0.53 GHz ($n$=0), can be excited.

In order to demonstrate this analytical prediction, we perform dynamical phase-field simulations (see Methods) to compute the spatiotemporal evolution of local magnetization in a 15-nm-thick (001) MAFO film upon the injection of a bipolar Gaussian acoustic pulse, which has a peak amplitude $\varepsilon_{max}$=0.3% in the adjacent (001) MAO and a duration $\tau$=6 ps, as indicated in Fig. 1a. Figure 2a presents the temporal profile of the $\Delta m_z(t)$ [=$m_z(t)$-$m_z(t$=0)] at the top surface ($z$=15 nm) of the MAFO film in first 0.5 ns after the acoustic pulse propagates into the film from its bottom surface ($z$=0 nm), which shows the features of mixed high-frequency magnon modes and their amplitudes gradually decrease due to magnetic damping. The evolution of the $\Delta m_z(t)$ at $t \geq$ 2.5 ns is plotted in Fig. 2b. At this stage, high-frequency magnon modes had attenuated, and a lower-frequency (ns-scale oscillation period), smaller-amplitude (~$10^{-5}$) magnon mode can be seen. Figure 2c shows the frequency spectra of both the $\Delta m_z(z$=15 nm,$t)$ within $t$=0-8.5 ns and the strain averaged over the MAFO thickness $<\varepsilon_{zz}>(t)$. As shown, three distinct magnon modes ($n$=0, 1 and 2) are excited and their frequency values match the analytical calculation almost exactly. Higher-mode magnons, e.g., 950 GHz for $n$=3 mode, were not excited, because their frequency values fall outside the frequency window of the acoustic pulse (0-600 GHz, shown in Fig. 2c). More detailed analyses indicate that the amplitudes of the magnons are proportional to the spectral amplitude of the acoustic pulse at the corresponding frequency (see Supplementary Note 1). For further demonstration, we extracted the spatial profiles of the three magnon modes by performing inverse Fourier transform of the spectrum $\Delta m_z(\omega)$ over the entire MAFO film with all non-peak-frequency components filtered out. As shown in Fig. 2d, the obtained profiles display canonical features of the $n$=0, 1, and 2 magnon modes (c.f., the schematics in Fig. 1a).

For detection, we suggest that the spectral features of these magnon modes should be retained in the frequency spectra of the charge current $\mathbf{J}^c$ in the adjacent Pt layer as well as the EM wave emitted by the $\mathbf{J}^c$. This is because $\mathbf{J}^c$ arise from the precessing local magnetization at the MAFO/Pt interface $\Delta\mathbf{m}(z$=15 nm,$t)$, and because the frequency spectrum of $\Delta\mathbf{m}(z$=15 nm,$t)$ (Fig. 2c) contains all the excited magnon modes. Figure 3a shows the spatiotemporal profile of the calculated total charge current density $J_x^c(z,t)$ in the Pt layer, which is a sum of the current generated via the iSHE $J_x^{iSHE}$ and the eddy current (polarization current) $J_x^p$ induced by the emitted EM wave (see details in Methods). As seen, the $J_x^c$ in the half thickness of the Pt layer is opposite to that in the other half. This is because (1) the directions of $\mathbf{J}^{iSHE}$ and $\mathbf{J}^p$ are opposite to each other; (2) the amplitude of $\mathbf{J}^{iSHE}$ decreases monotonically along the Pt thickness direction while $\mathbf{J}^p$ is almost spatially uniform in the Pt layer due to the long wavelength (millimeter-scale) of the emitted EM wave. Distributions of $J_x^{iSHE}$ and $J_x^p$ are shown in Supplementary Fig. S2. Figure 3b shows the temporal evolution of the electric-field component of the emitted EM wave $E_x(t)$ in the free space (at 5 nm above the Pt top surface), which decreases over time as the magnon amplitude decreases (c.f. Fig. 2a). The peak amplitude of the $E_x(t)$, ~80 V/m, is large enough for detection by the time-domain electro-optical sampling used in ultrafast THz emission spectroscopy (e.g., see refs.[22,23]). Figure 3c shows the frequency spectrum of the $E_x(t)$. Two discrete peak frequencies can be seen, which have the same values as those of $n$=1 and 2 mode magnon (c.f. Fig. 2c). Notably, although the spectral amplitude of the $n$=2 mode magnon is smaller than that of $n$=1 mode magnon (Fig. 2c), the spectral amplitude of the EM wave contributed by the $n$=2 mode magnon is larger because higher-frequency magnons pump larger-amplitude spin current into the Pt (see Methods). The EM wave emission from the $n=0$ mode magnon (FMR) is negligible because of its small amplitude (Fig. 2b) and low frequency (Fig. 2c). Furthermore, the emitted EM wave is circularly polarized, as shown in Fig. 3d, due to the phase difference between $E_x$ and $E_y$.

*Excitation and detection of sub-THz magnons in antiferromagnetic thin film*

We now show that coherent standing magnon modes can likewise be excited in an AFM thin film by a *ps* acoustic pulse via magnetoelastic coupling. A similar metal/dielectric/AFM/HM heterostructure is considered (Fig. 4a), using Al/MgO(001)/Fe$_{50}$Mn$_{50}$/Pt as an example. Polycrystalline Fe$_{50}$Mn$_{50}$ (FeMn) film is considered as the representative AFM material due to its robust magnetoelastic coupling ($B_1 \sim -9.7$ MJ m$^{-3}$)[33,34]. FeMn can be modeled as an easy-axis antiferromagnet with two magnetic sublattices[33,35] whose magnetizations are denoted as $\mathbf{m}^{(s)}$ ($s$ = 1,2). Following the same approach used for the FM film, an analytical formulation of the magnon dispersion relation $\omega(k)$ is derived for easy-axis AFM thin films (see Methods). Figure 4b shows the analytically calculated frequency $f$ ($=\omega/2\pi$) of each AFM magnon mode as a function of the FeMn film thickness $d$. Comparing Figs. 1b and Fig. 4b, it can be seen that the antiferromagnetic resonance (AFMR) frequency, or the $n=0$ mode AFM magnon, is much higher than the FMR frequency. If considering a 15-nm-thick FeMn film and the frequency window of the injected acoustic pulse reaches up to 200 GHz (illustrated as the vertical dashed line in Fig. 4b), three AFM magnon modes, with frequency of 192.6 GHz ($n=2$), 91.3 GHz ($n=1$), and 49.3 GHz ($n=0$), can be excited. The dynamical phase-field modeling results demonstrating this analytical prediction are shown in Fig. 5, where the evolution of both the $\Delta m_z^{(1)}$ and $\Delta m_z^{(2)}$ (see Fig. 5a) were utilized to analyze the frequency spectra of the AFM magnon modes (Fig. 5b). Moreover, as shown in Fig. 5a, the $\Delta m_z^{(1)}$ and $\Delta m_z^{(2)}$ are opposite to each other during the evolution, and it is noteworthy that the $\mathbf{m}^{(1)}$ and $\mathbf{m}^{(2)}$ are precessing counterclockwise and clockwise around the [111] easy axis, respectively, as sketched in the inset of Fig. 5a.

The precession of both the $\mathbf{m}^{(1)}$ and $\mathbf{m}^{(2)}$ at the FeMn/Pt interface can pump spin currents into the Pt layer. Principles of spin pumping from an easy-axis antiferromagnet have been discussed elsewhere[36–38]. Figure 6a shows the spatiotemporal profile of the total charge current density $J_x^c = J_x^{\text{iSHE}} + J_x^p$ in both FeMn and Pt films. The $J_x^c$ is larger in Pt because the $J_x^{\text{iSHE}}$ exists only in the Pt layer. Figure 6b presents the temporal evolution of the in-plane electric-field component $E_x(t)$ of the emitted EM wave in the free space (5 nm above the Pt top surface). The amplitude of $E_x(t)$ decreases over time as the magnon amplitude decreases (c.f., Fig. 5a). The frequency spectrum of the $E_x(t)$ in Fig. 6c reveals three distinct peak frequencies, which have same values as the excited AFM magnon modes $n=0$, 1 and 2 (c.f., Figs. 4b and 5b). Moreover, the emitted EM wave is linearly polarized with no phase difference between $E_x$ and $E_y$, as shown in Fig. 6d (see detailed explanation in Methods section).

**Discussion**

In summary, we used analytical calculations to show the physical principles of exciting standing magnons modes in both ferromagnetic (FM) and antiferromagnetic (AFM) thin films by ultrashort acoustic pulse. The frequencies of the high-order magnons can be sub-THz or higher depending on the film thickness (Fig. 1b and Fig. 4b). The analytical predictions were then demonstrated numerically by dynamical phase-field modeling. In the case of AFM thin films, the application of a bias magnetic field is not needed (c.f., Figs. 1a and 4a), which is an advantage for on-chip device integration due to the minimal crosstalk among neighboring units. We proposed that these acoustically excited sub-THz magnons can be detected by introducing an adjacent heavy metal (HM) thin film to enable spin pumping and spin-charge conversion. Specifically, since the frequency spectra of the magnon modes are similar to the frequency spectra of the charge current in the HM layer and the free-space electromagnetic (EM) wave emission, the excited magnons can be detected by measuring the EM wave emission. Compared to the commonly used method of ultrafast TR-MOKE where the signals of sub-THz magnons (which have nm-scale wavelength) can be averaged out due to the nm-scale penetration depth of the probe depths, detecting the EM wave emission allows for retaining the spectral information of all magnons modes.

In order to computationally demonstrate the proposed principles of detection and accurately model the emitted EM wave, we have developed an in-house dynamical phase-field model that considers fully coupled dynamics of acoustic waves, spin waves, and EM waves, which has previously not been considered together despite a few advanced computational models in this regard[18,39–41]. The physical validity and high numerical accuracy of our phase-field model can be seen from the almost exact match between the analytically calculated and simulated frequencies of the magnon modes (Figs. 2c and 5b). Results on the validation of other modules, especially our in-house finite-difference time-domain (FDTD) solver for EM wave generation and propagation, can be found in Supplementary Note 2 and Figs. S3-5). The predicted peak amplitude of the emitted electric field, on the order of 100 V/m, is sufficiently large for detection. The emission mainly arises from the charge current in the HM layer via electric dipole radiation, because the net magnetic dipole radiation from both high-order ($n\geq1$) standing magnons and the uniform magnetization precession ($n=0$) in nm-thick film are negligible — this also justifies the necessity of introducing the HM layer for magnon detection. Overall, this work computationally demonstrates a practical scheme to achieve the excitation and detection of exchange-dominated sub-THz coherent magnons, which has remained a challenge in the field of magnonics[30] despite its potential application towards high-speed magnonic devices. Looking ahead, the dynamical phase-field model we developed in this work can be utilized to accurately model ultrafast magnetoacoustics in more complex FM and AFM film-based heterostructures such as superlattices, as well as other physical processes and devices that involve phonon-magnon-photon coupling such as cavity magnonics[42,43] and mechanical antennas[44–47].

## Methods

*Dynamical phase-field model considering coupled dynamics of acoustic waves, spin waves, and electromagnetic waves in ferromagnetic and antiferromagnetic heterostructures*

Below we describe the individual modules for spin wave (magnon) dynamics, acoustic wave (coherent phonons), and EM wave (photon) dynamics, and how these modules are coupled together in a multilayer heterostructure. Here the "coupled" means bidirectional phonon-magnon and magnon-photon coupling. Specifically, the acoustically excited local magnetization will generate secondary acoustic waves via magnetoelastic backaction. The EM wave, which originates from the precession of magnetic dipoles (via spin-charge conversion), will in turn affect the magnetization dynamics via its magnetic-field component. Both types of back-actions, despite being weak in the present cases, have been incorporated. Moreover, our model is GPU (Graphics Processing Unit) accelerated to facilitate high-throughput modeling and heterostructure design.

Part 1: Spin wave dynamics by solving the Landau-Lifshitz-Gilbert (LLG) equation

The simulations were performed with a dynamical phase-field model that considers the fully coupled dynamics of elastic waves (acoustic phonons), spin waves (magnons), and EM waves in a heterostructure with discontinuous magnetic and elastic properties across the interface. The evolution of normalized local magnetization **m** in a ferromagnetic (FM) system and $\mathbf{m}^{(s)}$ ($s=1,2$) in two magnetic sublattices of an antiferromagnetic (AFM) system are governed by the LLG equation. The LLG equation for the FM system is expresses as,

$$\frac{\partial \mathbf{m}}{\partial t} = -\frac{\gamma}{1+\alpha^2} \mathbf{m} \times \mathbf{H}_{\text{eff}} - \frac{\alpha\gamma}{1+\alpha^2} \mathbf{m} \times (\mathbf{m} \times \mathbf{H}_{\text{eff}}) \qquad (1)$$

where $\gamma$ is the gyromagnetic ratio; $\alpha$ is the magnetic damping coefficient. The total effective magnetic field $\mathbf{H}_{\text{eff}} = \mathbf{H}^{\text{anis}} + \mathbf{H}^{\text{exch}} + \mathbf{H}^{\text{mel}} + \mathbf{H}^{\text{dip}} + \mathbf{H}^{\text{ext}} + \mathbf{H}^{\text{EM}}$. Among them, the magnetocrystalline anisotropy field $\mathbf{H}^{\text{anis}}$ is given by,

$$H_i^{\text{anis}} = -\frac{2}{\mu_0 M_s}\left[K_1\left(m_j^2 + m_k^2\right) + K_2 m_j^2 m_k^2\right] m_i, \qquad (2)$$

where $\mu_0$ is vacuum permeability; $M_s$ is saturation magnetizatio; $K_1$ and $K_2$ are magnetocrystalline anisotropy coefficients; $i = x, y, z$, and $j \neq i$, $k \neq i, j$. The magnetic exchange coupling field $\mathbf{H}^{\text{exch}} = \frac{2A_{\text{ex}}}{\mu_0 M_s}\nabla^2 \mathbf{m}$. The magnetic boundary condition[48] $\partial \mathbf{m}/\partial z = 0$ is applied on the surfaces of the magnetic film. The magnetoelastic field $\mathbf{H}^{\text{mel}}$ describes the influence of elastic strain $\boldsymbol{\varepsilon}$ on the local magnetization dynamics, written as,

$$H_i^{\text{mel}} = -\frac{2}{\mu_0 M_s}\left[B_1 m_i \varepsilon_{ii} + B_2\left(m_j \varepsilon_{ij} + m_k \varepsilon_{ik}\right)\right], \qquad (3)$$

where $B_1$ and $B_2$ are magnetoelastic coupling coefficients; $i = x, y, z$, and $j \neq i$, $k \neq i, j$. The magnetic dipolar coupling field $\mathbf{H}^{\text{dip}} = (0, 0, -M_s m_z)$ for an infinitely large $xy$ plane within which the magnetization **m** is spatially uniform. The external bias magnetic field $\mathbf{H}^{\text{ext}} = (0, 0, H_z^{\text{ext}})$ is applied along the $z$ axis to lift magnetizations off the $xy$ plane by 45° before acoustic excitation, so that the torque exerted by the $\mathbf{H}^{\text{mel}}$ on the magnetizations is maximized. The magnetic field component of the emitted EM wave $\mathbf{H}^{\text{EM}}$ describes the back-action of the EM wave on local magnetization dynamics. The calculation of $\mathbf{H}^{\text{EM}}$ will be detailed later.

The LLG equations for the two magnetic sublattices of the AFM system have the same form as the FM but with their own total effective field $\mathbf{H}_{\text{eff}}^{(s)}$ ($s=1,2$),

$$\frac{\partial \mathbf{m}^{(s)}}{\partial t} = -\frac{\gamma}{1+\alpha^2}\mathbf{m}^{(s)}\times\mathbf{H}_{\text{eff}}^{(s)} - \frac{\alpha\gamma}{1+\alpha^2}\mathbf{m}^{(s)}\times\left(\mathbf{m}^{(s)}\times\mathbf{H}_{\text{eff}}^{(s)}\right). \tag{4}$$

In this work, the gyromagnetic ratio $\gamma$ and magnetic damping coefficient $\alpha$ are set to be same for both sublattices as approximation, following ref.[33,35]. The total effective magnetic field $\mathbf{H}_{\text{eff}}^{(s)}=\mathbf{H}^{\text{anis},(s)}+\mathbf{H}^{\text{exch},(s)}+\mathbf{H}^{\text{mel},(s)}+\mathbf{H}^{\text{AFM},(s)}+\mathbf{H}^{\text{dip}}+\mathbf{H}^{\text{ext}}+\mathbf{H}^{\text{EM}}$. The model considers an easy-axis AFM system for which the magnetocrystalline anisotropy fields $\mathbf{H}^{\text{anis},(s)}$ in both sublattices are given by,

$$\mathbf{H}^{\text{anis},(s)}=\frac{2K_{u1}}{\mu_0 M_s}[\mathbf{a}_u\cdot\mathbf{m}^{(s)}]\mathbf{a}_u, \tag{5}$$

where $K_{u1}$ is uniaxial anisotropy coefficient; $\mathbf{a}_u$ is unit vector in the direction of easy axis; $K_{u1}$ and $M_s$ are assumed to be same for the two sublattices. For illustration, $\mathbf{a}_u$ is set to be along [111] in this work and the $\mathbf{m}^{(1)}$ and $\mathbf{m}^{(2)}$ are directed along [111] and [$\bar{1}\bar{1}\bar{1}$] before acoustic excitation, respectively. The principles of magnon excitation and detection are independent of the easy-axis orientation. The intra-lattice magnetic exchange coupling field $\mathbf{H}^{\text{exch},(s)}=\frac{2A_{\text{ex}}}{\mu_0 M_s}\nabla^2\mathbf{m}^{(s)}$ has the same form as that in the FM system and $A_{\text{ex}}$ is assumed to be same for two sublattices. The magnetic boundary condition $\partial\mathbf{m}^{(s)}/\partial z=0$ is applied on the two surfaces of the AFM film[49]. The magnetoelastic field is halved in magnitude for AFM in comparison to the form for FM (Eq. 3) to ensure that saturation magnetostriction $\lambda_s$ occurs when $\mathbf{m}^{(1)}$ and $\mathbf{m}^{(2)}$ are coaxially directed[33],

$$H_i^{\text{mel},(s)}=-\frac{1}{\mu_0 M_s}\left[B_1 m_i^{(s)}\varepsilon_{ii}+B_2\left(m_j^{(s)}\varepsilon_{ij}+m_k^{(s)}\varepsilon_{ik}\right)\right], \tag{6}$$

where magnetoelastic coupling effects in both sublattices are assumed to be same. Therefore, the same set of $B_1$ and $B_2$ is used for both sublattices; $i = x, y, z,$ and $j \neq i, k \neq i, j$. It is noteworthy that $B_1=B_2$ for magnets with isotropic elasticity. The two sublattices also share the same magnetic dipolar coupling field $\mathbf{H}^{\text{dip}}=(0, 0, -M_s m_z^{(1)}-M_s m_z^{(2)})$ which includes contribution from both sublattices. Moreover, we set $\mathbf{H}^{\text{ext}}=0$. Before acoustic excitation, the AFM film is set to be a single AFM domain, that is, the Néel vector $\mathbf{n}=0.5[\mathbf{m}^{(1)}-\mathbf{m}^{(2)}]$ is spatially uniform and along the [111] easy axis. Such single AFM domain could be obtained by cooling the magnet from its high-temperature paramagnetic phase in the presence of a bias magnetic field applied along the easy axis. After then, the bias magnetic field can be removed. Different from the case of FM film, a single AFM domain can remain stable at $\mathbf{H}^{\text{ext}}=0$. The effective field produced by the inter-lattice AFM-type exchange coupling[33] favors opposite alignment of $\mathbf{m}^{(1)}$ and $\mathbf{m}^{(2)}$, calculated as,

$$\mathbf{H}^{\text{AFM},(1)} = -\frac{J}{\mu_0 M_s}\mathbf{m}^{(2)}, \tag{7a}$$

$$\mathbf{H}^{\text{AFM},(2)} = -\frac{J}{\mu_0 M_s}\mathbf{m}^{(1)}, \tag{7b}$$

where $J$ is the AFM exchange coupling coefficient. The sublattices share the same $\mathbf{H}^{\text{EM}}$ from EM wave.

Part 2: EM dynamics in magnet/heavy-metal heterostructure by solving Maxwell's equations

To calculate the EM wave dynamics of electric and magnetic dipole radiation, an in-house FDTD solver of Maxwell's equations was developed. The two governing equations are listed below.

$$\nabla \times \mathbf{E}^{EM} = -\mu_0 \left( \frac{\partial \mathbf{H}^{EM}}{\partial t} + \frac{\partial \mathbf{M}}{\partial t} \right), \tag{8}$$

$$\nabla \times \mathbf{H}^{EM} = \frac{\partial \mathbf{D}}{\partial t} + \mathbf{J}^p + \mathbf{J}^f, \tag{9}$$

where $\mathbf{E}^{EM}$ is the electric field component of the EM wave for the main results; $\mathbf{M}$ is the local magnetization. The temporally changing $\mathbf{M}$ provides the source of magnetic dipole radiation. For FM system, $\mathbf{M}=M_s\mathbf{m}$, where the $\mathbf{m}$ can be obtained by solving the LLG equation (Eq. 1). For AFM system, $\mathbf{M}=\mathbf{M}^{(1)}+\mathbf{M}^{(2)}=M_s[\mathbf{m}^{(1)}+\mathbf{m}^{(2)}]$ (assuming that $M_s$ is the same in both sublattices), where $\mathbf{m}^{(1)}$ and $\mathbf{m}^{(2)}$ are likewise obtained by solving LLG equations (Eq. 4). The LLG and Maxwell's equations (Eqs. 8 and 9) are coupled and solved simultaneously. $\mathbf{D}$ is electric displacement field; $\mathbf{J}^p$ is polarization current induced by the electric field $\mathbf{E}^{EM}$ in dispersive medium, which causes the absorption and reflection of the EM wave. The polarization current $\mathbf{J}^p$ (or eddy current) in metallic conductors (Pt and FeMn in this work) is obtained by solving time-domain auxiliary differential equation (ADE) based on Drude model[50],

$$\frac{\partial \mathbf{J}^p}{\partial t} + \tau^{-1}\mathbf{J}^p = \varepsilon_0 \omega_p^2 \mathbf{E}^{EM}, \tag{10}$$

where $\omega_p$ and $\tau$ denote the plasma frequency and electron relaxation time, respectively. For alloy such as $Fe_{50}Mn_{50}$, the total $\mathbf{J}^p$ is treated as the composition-weighted average of the $\mathbf{J}^p$ of each metallic component, i.e., $0.5\mathbf{J}^p_{Fe} + 0.5\mathbf{J}^p_{Mn}$. This ensures that the frequency-dependent relative permittivity $\varepsilon_r(\omega)$ of the alloy is also the composition-weighted average of the $\varepsilon_r(\omega)$ of constituent metallic components[51]. The ADE (Eq. 10) and the Maxwell's equations (Eqs. 8 and 9) are coupled and solved simultaneously. The $\mathbf{J}^f$ is the free charge current density. In this work, $\mathbf{J}^f$ is converted from the spin current density $\mathbf{J}^s$ via the iSHE ($\mathbf{J}^f=\mathbf{J}^{iSHE}$) and is the source of electric dipole radiation.

The spin current density $\mathbf{J}_0^s(t) = \mathbf{J}^s(z=d, t)$ at the FM/Pt interface is evaluated via the relation[52] $\mathbf{e}_n \cdot \mathbf{J}_0^s = \frac{\hbar}{4\pi}\text{Re}[g_{eff}^{\uparrow\downarrow}]\mathbf{m}\times\frac{\partial \mathbf{m}}{\partial t}$, where $\mathbf{e}_n$ is the unit vector normal to the FM/Pt interface and pointing to Pt, $\hbar$ is the reduced Planck constant, and $\text{Re}[g_{eff}^{\uparrow\downarrow}]$ is the real part of effective spin-mixing conductance; $\mathbf{m}$ is obtained by solving the LLG equation (Eq. 1). The spin current density $\mathbf{J}_0^s(t)$ from the AFM spin pumping is calculated as the sum of contribution from both sublattices via the relation[36], $\mathbf{e}_n \cdot \mathbf{J}_0^s = \frac{\hbar}{4\pi}\text{Re}[g_{eff}^{\uparrow\downarrow}](\mathbf{m}^{(1)}\times\frac{\partial \mathbf{m}^{(1)}}{\partial t}+\mathbf{m}^{(2)}\times\frac{\partial \mathbf{m}^{(2)}}{\partial t})$, where $\text{Re}[g_{eff}^{\uparrow\downarrow}]$ is assumed to be the same in both sublattices. Due to the dissipative propagation of the spin accumulation in the heavy metal, the resultant spin current decays as a function of the distance from the FM (or AFM)/Pt interface[53], $\mathbf{J}^s(z,t)=\mathbf{J}_0^s(t)\frac{\sinh[(d+d_{Pt}-z)/\lambda_{sd}]}{\sinh(d_{Pt}/\lambda_{sd})}$, where $d_{Pt} = 7$ nm is the thickness of the Pt layer, $\lambda_{sd}$ is the spin diffusion length in Pt. The iSHE charge current density[52] $\mathbf{J}^{iSHE}(z,t)=\theta_{Pt}\frac{2e}{\hbar}\mathbf{e}_n\times[\mathbf{e}_n\cdot\mathbf{J}^s(z,t)]$ in the Pt layer, where $\theta_{Pt}$ is the spin Hall angle of Pt and $e$ is elementary charge. The formula of $\mathbf{J}^{iSHE}$ indicates that it has only in-plane components ($J_z^{iSHE}=0$). Therefore, both the magnetic and electric-fields of the emitted plane EM wave only have in-plane components, that is, $E_z^{EM}=0$ and $H_z^{EM}=0$.

The type of polarization (circular or linear) for the emitted EM wave depends on the total $\mathbf{J}^{iSHE}$ in the Pt layer. In the case of FM/Pt, the $\mathbf{J}^{iSHE}$ is circularly polarized due to phase difference between different components of Δ**m**. In the case of AFM/Pt, although the iSHE current produced by each magnetic sublattice **J**$^{\text{iSHE},(s)}$ are circularly polarized, the total **J**$^{\text{iSHE}}$=**J**$^{\text{iSHE},(1)}$+**J**$^{\text{iSHE},(2)}$ is linearly polarized, giving rise to linear-polarized EM wave (see Fig. 6d). To understand this, we use $e^{-i\omega t}$ and $e^{i(\omega t+\varphi)}$ to describe the **J**$^{\text{iSHE},(1)}$ and **J**$^{\text{iSHE},(2)}$ that have opposite chirality due to the counterclockwise and clockwise magnetization precession, respectively, where $\varphi$ is their phase difference. The real (Re) and imaginary (Im) part of the $e^{-i\omega t}$ represents the $x$ and $y$ component of the **J**$^{\text{iSHE},(1)}$, and likewise for the $e^{i(\omega t+\varphi)}$. Therefore, the $J_y^{\text{iSHE}}/J_x^{\text{iSHE}}$ for the **J**$^{\text{iSHE}}$ can be calculated as Im$[e^{-i\omega t}+e^{i(\omega t+\varphi)}]$/Re$[e^{-i\omega t}+e^{i(\omega t+\varphi)}]$ which gives a time-independent value $\tan(\varphi/2)$. This indicates a linear polarization for the **J**$^{\text{iSHE}}$.

Part 3: Acoustic wave dynamics by solving the elastodynamic equation

The evolution of mechanical displacement **u** is obtained by solving an elastodynamic equation which incorporates the magnetostrictive stress to describe the backaction of magnetization dynamics on elastic wave dynamics, $\rho \frac{\partial^2 \mathbf{u}}{\partial t^2}=\nabla\cdot[\mathbf{c}(\boldsymbol{\varepsilon}-\boldsymbol{\varepsilon}^0)]$, where $\rho$ and **c** are phase-dependent mass density and elastic stiffness, respectively. A full tensorial expansion of this elastodynamic equation is provided in our previous work[19]. The strain $\boldsymbol{\varepsilon}$ is related to **u** via $\varepsilon_{ij}=\frac{1}{2}\left(\frac{\partial u_i}{\partial j}+\frac{\partial u_j}{\partial i}\right)$; $i,j=x,y,z$. The $\boldsymbol{\varepsilon}^0$ is eigenstrain produced by magnetization via magnetostriction. For a FM system with a cubic parent phase, $\boldsymbol{\varepsilon}^0$ takes the conventional form of,

$$\varepsilon_{ii}^0=\frac{3}{2}\lambda_{100}^{\text{FM}}\left(m_i^2-\frac{1}{3}\right), \varepsilon_{ij}^0=\frac{3}{2}\lambda_{111}^{\text{FM}}m_i m_j, \qquad (11)$$

where $\lambda_{100}^{\text{FM}}$ and $\lambda_{111}^{\text{FM}}$ are saturation magnetostriction along the local <100> and <111> axes. For an isotropic AFM system, one has[33],

$$\varepsilon_{ii}^0=\frac{3}{4}\lambda_s^{\text{AFM}}\left(m_i^{(1)2}-\frac{1}{3}\right)+\frac{3}{4}\lambda_s^{\text{AFM}}\left(m_i^{(2)2}-\frac{1}{3}\right), \varepsilon_{ij}^0=\frac{3}{4}\lambda_s^{\text{AFM}}m_i^{(1)}m_j^{(1)}+\frac{3}{4}\lambda_s^{\text{AFM}}m_i^{(2)}m_j^{(2)}, \qquad (12)$$

where $\lambda_s^{\text{AFM}}$ is the magnetostrictive coefficient. Note that $i=x,y,z$, $j \neq i$. The LLG equations (Eqs. 1 and 4) and the elastodynamic equation are coupled and solved simultaneously. For simplicity, the injection of the ultrashort acoustic pulse is modelled by applying a time-dependent mechanical displacement at the Al/MAO or Al/MgO interface in the form of Gaussian function[18], $u_z(t)=u_{\max}\exp(-t^2/\sigma^2)$, which leads to a bipolar longitudinal strain $\varepsilon_{zz}=\partial u_z/\partial z$ propagating in the dielectric layer (MAO or MgO) and then the magnetic layer (MAFO or FeMn), as sketched in Fig. 1a and 4a. Such bipolar Gaussian acoustic pulse is a commonly used approximation for describing fs-laser-induced ultrafast acoustic pulse from polycrystalline metal transducer such as Al[7,9,14]. The dielectric layer (MAO or MgO) is set to be sufficiently thick (e.g., hundreds of micrometers) to serve as a perfect sink of the reflected acoustic pulse. The entire heterostructure is discretized into one-dimensional (1D) computational cells along $z$ direction, with cell size $\Delta z$=0.2 nm. Central finite difference is used for calculating spatial derivatives. All equations are solved simultaneously using the classical Runge-Kutta method for time-marching with a real-time step $\Delta t = 5\times10^{-19}$ s.

The materials parameters are summarized below. For (001) MAO[54], the elastic stiffness coefficients $c_{11}$ = 282.9 GPa, $c_{12}$ = 155.4 GPa, $c_{44}$ = 154.8 GPa and mass density $\rho$ = 3578 kg m$^{-3}$. For (001) MAFO thin film[29,55], the elastic stiffness coefficients are assumed to be the same as MAO; $\rho$ = 4355 kg m$^{-3}$; gyromagnetic ratio $\gamma$ = 0.227 rad MHz A$^{-1}$ m; the damping coefficient $\alpha$ = $\alpha^0$+ $\alpha^s$,[56] where $\alpha^0$ = 0.0015 is the intrinsic Gilbert damping coefficient without spin pumping;

$\alpha^s = \frac{g\mu_B}{4\pi M_s} g_{eff}^{\uparrow\downarrow} \frac{1}{d}$ is the magnetic damping induced by spin pumping[38] ($g = 2.05$ is the g-factor[29], $\mu_B$ is the Bohr magneton); saturation magnetization $M_s = 0.0955$ MA m$^{-1}$; the exchange coupling coefficient $A_{ex} = 4$ pJ m$^{-1}$ is assumed to be same as CoFe$_2$O$_4$[57], magnetocrystalline anisotropy coefficient $K_1 = -477.5$ J m$^{-3}$; magnetoelastic coupling coefficient $B_1 = 1.2$ MJ m$^{-3}$ and $B_2 = 0$. For MgO[58], $c_{11} = 297$ GPa, $c_{12} = 95.9$ GPa, $c_{44} = 156$ GPa and $\rho = 3580$ kg m$^{-3}$. For polycrystalline FeMn with isotropic elastic properties[33], $c_{11} = 103.7$ GPa, $c_{12} = 44.4$ GPa, $c_{44} = 29.6$ GPa, which are calculated based on Young's modulus of 77 GPa and Poisson's ratio of 0.3 and assumption of isotropic elasticity; $\rho = 7700$ kg m$^{-3}$; $\gamma = 0.221$ rad MHz A$^{-1}$ m; the intrinsic Gilbert damping coefficient $\alpha^0 = 0.0045$; $M_s = 1.7$ MA m$^{-1}$; $A_{ex} = 21$ pJ m$^{-1}$ (some of these parameters are assumed to be same as the Fe[59–61] for simplicity). The uniaxial anisotropy coefficient $K_{u1}$ is assumed to be $5 \times 10^5$ J m$^{-3}$. The coefficient for AFM-type exchange coupling $J = 3.97 \times 10^6$ J m$^{-3}$. The magnetostrictive coefficients $\lambda_s^{AFM} = 109$ ppm based on a recent experiment[34], hence $B_1 = B_2 = -1.5\lambda_s^{AFM}(c_{11}-c_{12}) = -3c_{44}\lambda_s^{AFM} = -9.7$ MJ m$^{-3}$. For Pt[62], $c_{11} = 347$ GPa, $c_{12} = 250$ GPa, $c_{44} = 75$ GPa and $\rho = 21450$ kg m$^{-3}$. The plasma frequency $\omega_p$ and electron relaxation time $\tau$ for Pt, Fe and Mn can be found in ref. [63]. For Pt, $\omega_p = 9.1$ rad fs$^{-1}$ and $\tau = 7.5$ fs; for Fe, $\omega_p = 6.4265$ rad fs$^{-1}$ and $\tau = 12$ fs; for Mn, $\omega_p = 8.1737$ rad fs$^{-1}$ and $\tau = 0.9$ fs. For the case of MAFO/Pt[31], the effective spin-mixing conductance $g_{eff}^{\uparrow\downarrow} = 3.36 \times 10^{18}$ m$^{-2}$, the spin diffusion length in Pt $\lambda_{sd} = 3.3$ nm, and the spin Hall angle of Pt $\theta_{Pt} = 0.83$; these parameters for the case of FeMn/Pt are assumed to be same as those for Fe/Pt[32,64], which are $g_{eff}^{\uparrow\downarrow} = 4.9 \times 10^{19}$ m$^{-2}$, $\lambda_{sd} = 3.4$ nm, $\theta_{Pt} = 0.056$.

*Magnon dispersion relation for FM and AFM thin films*

The magnon dispersion relation $\omega(k)$ can be obtained analytically from the linearization of the LLG equation under zero magnetic damping ($\alpha = 0$). Detailed procedures are shown in Supplementary Note 3. For FM thin films,

$$\omega(k) = \sqrt{-(A_{32}A_{23} + A_{12}A_{21} + A_{13}A_{31})}, \quad (13)$$

where

$$A_{12} = \gamma\left[Dk^2 m_z^0 + H_z^{ext} - M_s m_z^0 - \frac{2K_1}{\mu_0 M_s}\left(3m_y^{0^2} m_z^0 - m_z^{0^3}\right)\right], \quad (14)$$

$$A_{13} = \gamma\left[-Dk^2 m_y^0 - M_s m_y^0 - \frac{2K_1}{\mu_0 M_s}\left(m_y^{0^3} - 3m_y^0 m_z^{0^2}\right)\right], \quad (15)$$

$$A_{21} = \gamma\left[-Dk^2 m_z^0 - H_z^{ext} + M_s m_z^0 - \frac{2K_1}{\mu_0 M_s}\left(m_z^{0^3} - 3m_z^0 m_x^{0^2}\right)\right], \quad (16)$$

$$A_{23} = \gamma\left[Dk^2 m_x^0 + M_s m_x^0 - \frac{2K_1}{\mu_0 M_s}\left(3m_z^{0^2} m_x^0 - m_x^{0^3}\right)\right], \quad (17)$$

$$A_{31} = \gamma\left[Dk^2 m_y^0 - \frac{2K_1}{\mu_0 M_s}\left(3m_x^{0^2} m_y^0 - m_y^{0^3}\right)\right], \quad (18)$$

$$A_{32} = \gamma\left[-Dk^2 m_x^0 - \frac{2K_1}{\mu_0 M_s}\left(m_x^{0^3} - 3m_x^0 m_y^{0^2}\right)\right]. \quad (19)$$

with the exchange stiffness $D = \frac{2A_\text{ex}}{\mu_0 M_\text{s}}$ and magnetization vectors at equilibrium $\mathbf{m}^0 = (m_x^0, m_y^0, m_z^0)$. For AFM thin films,

$$\omega = \frac{2\gamma}{\mu_0 M_\text{s}} \sqrt{(K_\text{u1} + A_\text{ex} k^2)(K_\text{u1} + A_\text{ex} k^2 + J)}. \tag{20}$$

**Data Availability**

The data that support the plots presented in this paper and its supplemental information files are available from the corresponding author upon reasonable request.

**Code Availability**

The codes for the dynamical phase-field model with coupled magnon-phonon-EM wave dynamics are available from the corresponding author upon reasonable request.

**Acknowledgements**

J.-M.H. acknowledges support from the NSF award CBET-2006028 and the Accelerator Program from the Wisconsin Alumni Research Foundation. The simulations were performed using Bridges at the Pittsburgh Supercomputing Center through allocation TG-DMR180076, which is part of the Extreme Science and Engineering Discovery Environment (XSEDE) and supported by NSF grant ACI-1548562.

**Author Contributions**

J.-M.H. conceived the idea, designed and supervised the research. S.Z. derived the analytical formulae, developed the computer codes of the dynamical phase-field model, and performed the research. Both authors analyzed the data and wrote the paper.

**Competing interests**

The authors declare no competing interests.

**Figures and Captions**

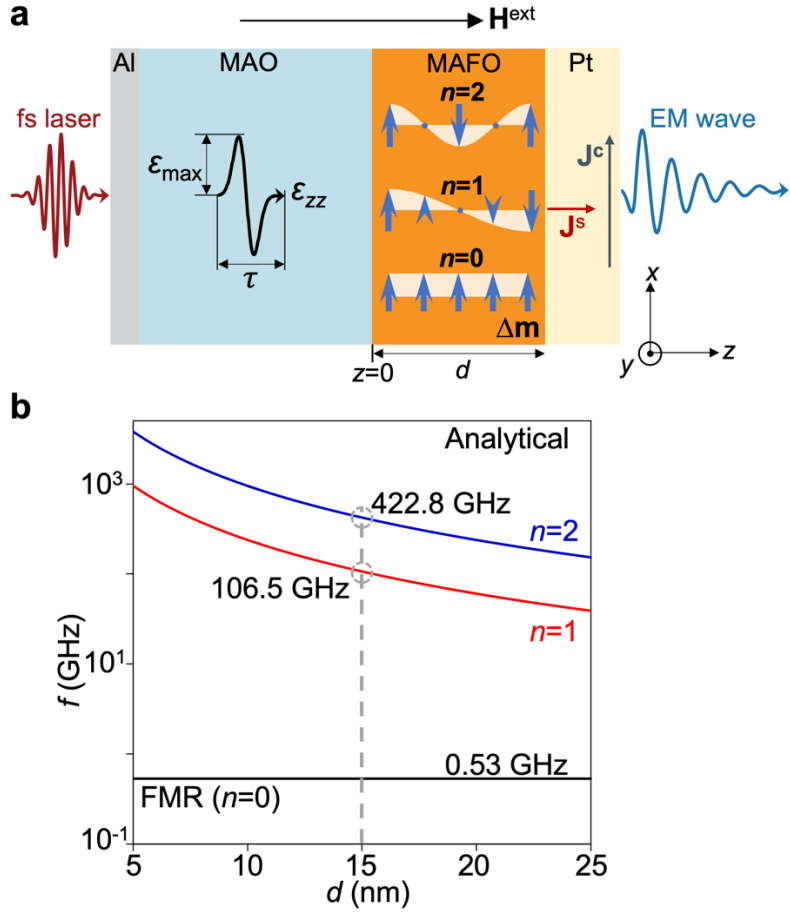

**Figure 1**. **Ultrafast magnetoacoustics in FM thin films.** (**a**) Schematic of Al/MAO/MAFO/Pt heterostructure as an example for acoustic excitation of sub-THz magnons in a FM thin film and their detection via electromagnetic (EM) wave emission. The magnons (denoted by Δ**m**) of different modes *n* are excited in the MAFO layer by fs-laser-induced ultrashort acoustic pulse $\varepsilon_{zz}$ which has a peak amplitude $\varepsilon_{max}$ and a duration $\tau$. The excited magnons pump spin current **J**$^s$ into the Pt layer, which is converted to transverse charge current **J**$^c$ via iSHE. **J**$^c$ emits EM wave via electric dipole radiation. (**b**) Analytically calculated frequencies of the *n*=1 and *n*=2 mode magnons and the ferromagnetic resonance (FMR) mode (*n*=0) magnon as a function of MAFO film thickness (*d*). The frequencies for the case of *d*=15 nm are labeled.

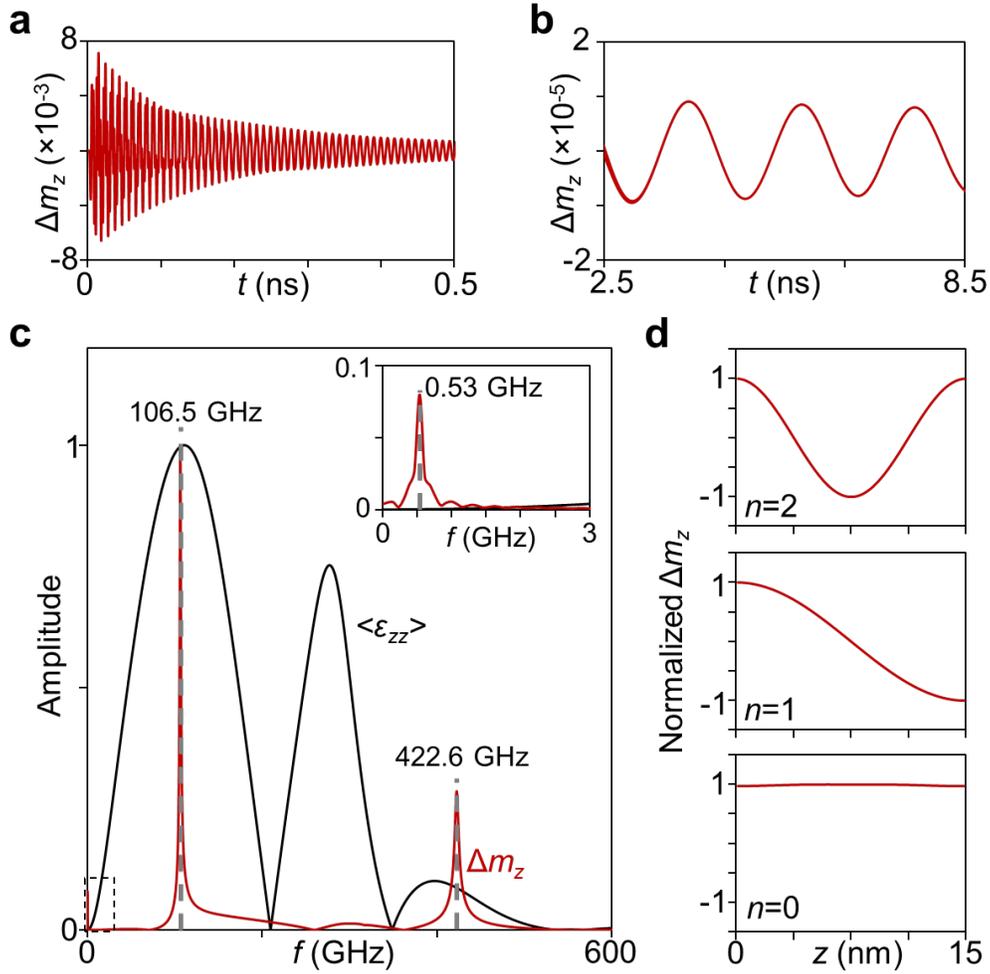

**Figure 2. Magnon excitation.** Temporal profiles of the $\Delta m_z(z=d,t)$ at the top surface of the MAFO within (**a**) $t$=0-0.5 ns showing the evolution of higher-mode ($n\geq1$) exchange magnon and (**b**) $t$ = 2.5-8.5 ns showing the low-frequency precession of $n$=0 mode magnon. Note that $\Delta m_z(z,t) = m_z(z,t)-m_z(z,t=0)$, and $t = 0$ is defined as the moment the acoustic pulse enters the MAFO film from its bottom surface ($z$=0). (**c**) Frequency spectra of the $\Delta m_z(z=d,t)$ within $t$= 0-8.5 ns (red line) and the thickness-averaged acoustic pulse $<\varepsilon_{zz}>(t)$ (black line). Inset: enlarged section within frequency $f$ = 0-3 GHz. (**d**) Spatial profiles of the $n$ = 2, 1 and 0 mode magnon components, represented by the normalized $\Delta m_z(z,t)$ across the 15-nm-thick MAFO.

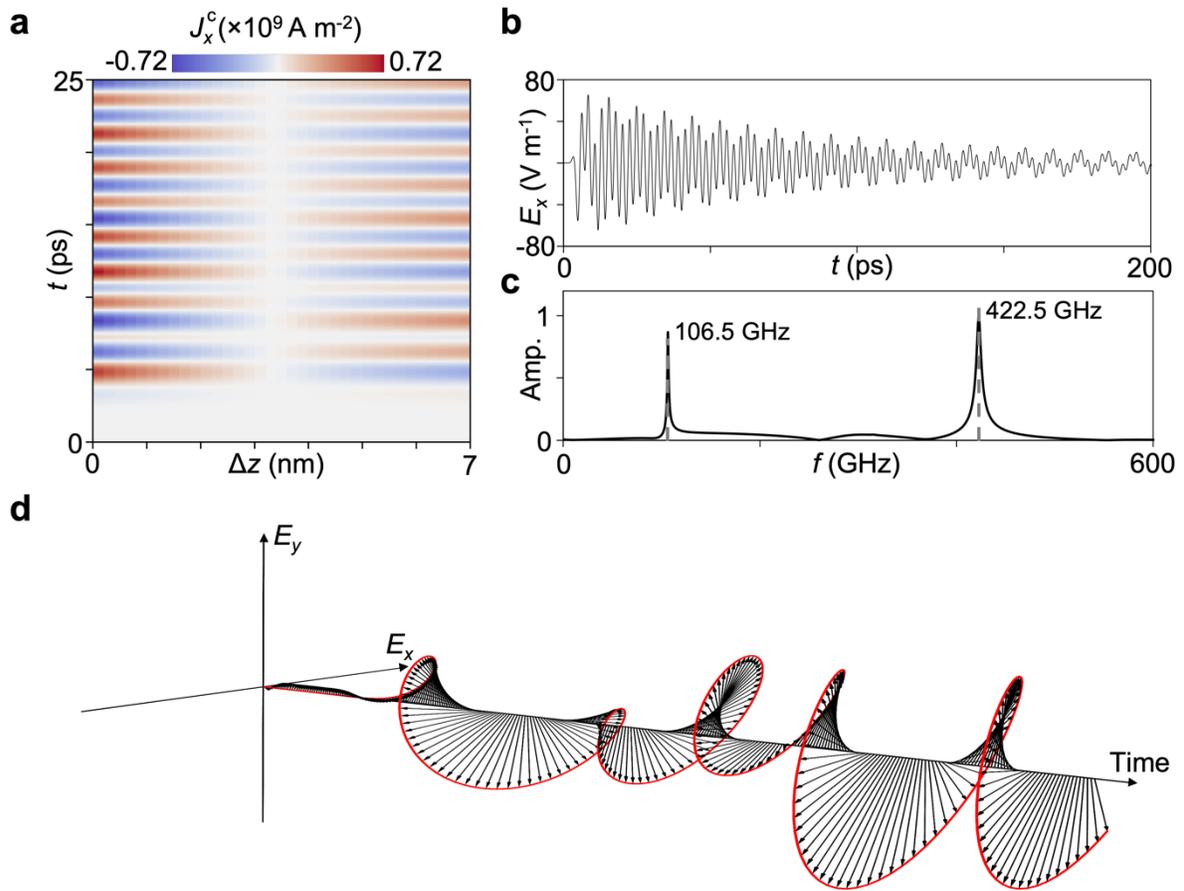

**Figure 3. Magnon detection.** (**a**) Spatiotemporal profile of the total charge current density $J_x^c$ in the 7-nm-thick Pt. (**b**) Temporal profile of the emitted electric-field component $E_x(t)$ at 5 nm above the Pt top surface within $t$ = 0-200ps and (**c**) its frequency spectrum. (**d**) Evolution of the emitted electric field vector $\mathbf{E}(t)$ within $t$=0-15 ps, showing a circular polarization.

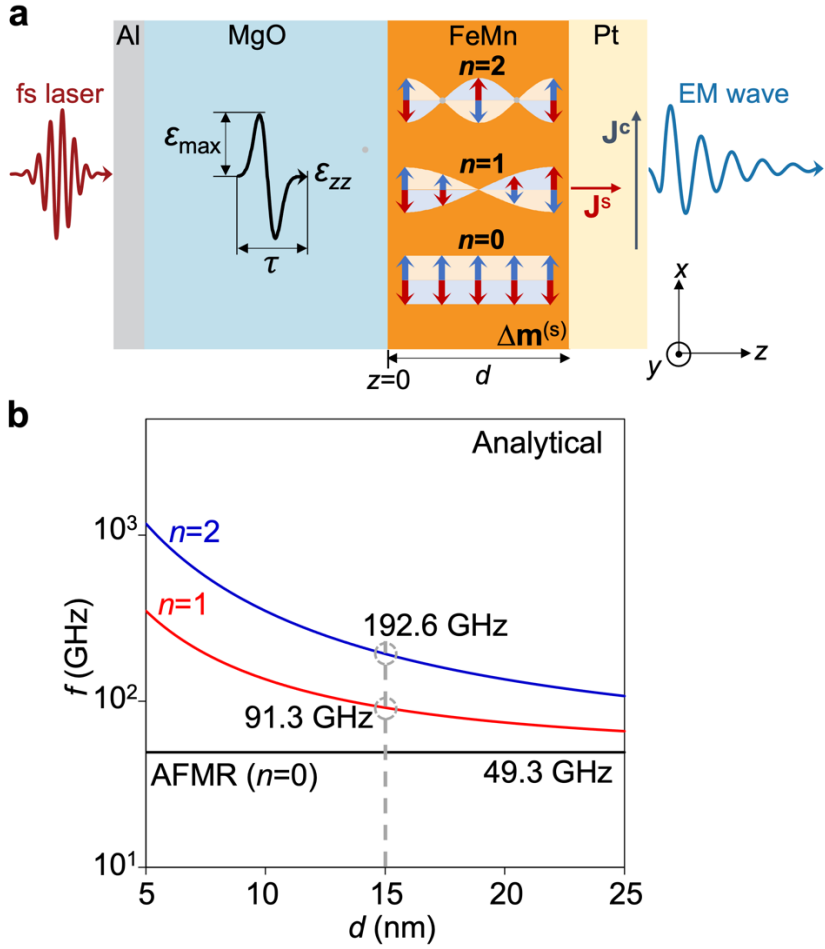

**Figure 4. Ultrafast magnetoacoustics in AFM thin films.** (**a**) Schematic of Al/MgO/Fe$_{50}$Mn$_{50}$/Pt heterostructure as an example for acoustic excitation of sub-THz magnons in an AFM thin film and their detection via electromagnetic (EM) wave emission. In both magnetic sublattices of the Fe$_{50}$Mn$_{50}$ layer, the magnons (denoted by $\Delta \mathbf{m}^{(s)}$) of different modes $n$ are excited by fs-laser-induced ultrashort acoustic pulse $\varepsilon_{zz}$ which has a peak amplitude $\varepsilon_{max}$ and a duration $\tau$. The excited magnons pump spin current $\mathbf{J}^s$ into the Pt layer, which is converted to transverse charge current $\mathbf{J}^c$ via iSHE. $\mathbf{J}^c$ emits EM wave via electric dipole radiation. (**b**) Analytically calculated frequencies of the $n=1$ and $n=2$ mode magnons and the antiferromagnetic (AFMR) mode ($n=0$) magnon as a function of FeMn film thickness ($d$). The frequencies for the case of $d=15$ nm are labeled.

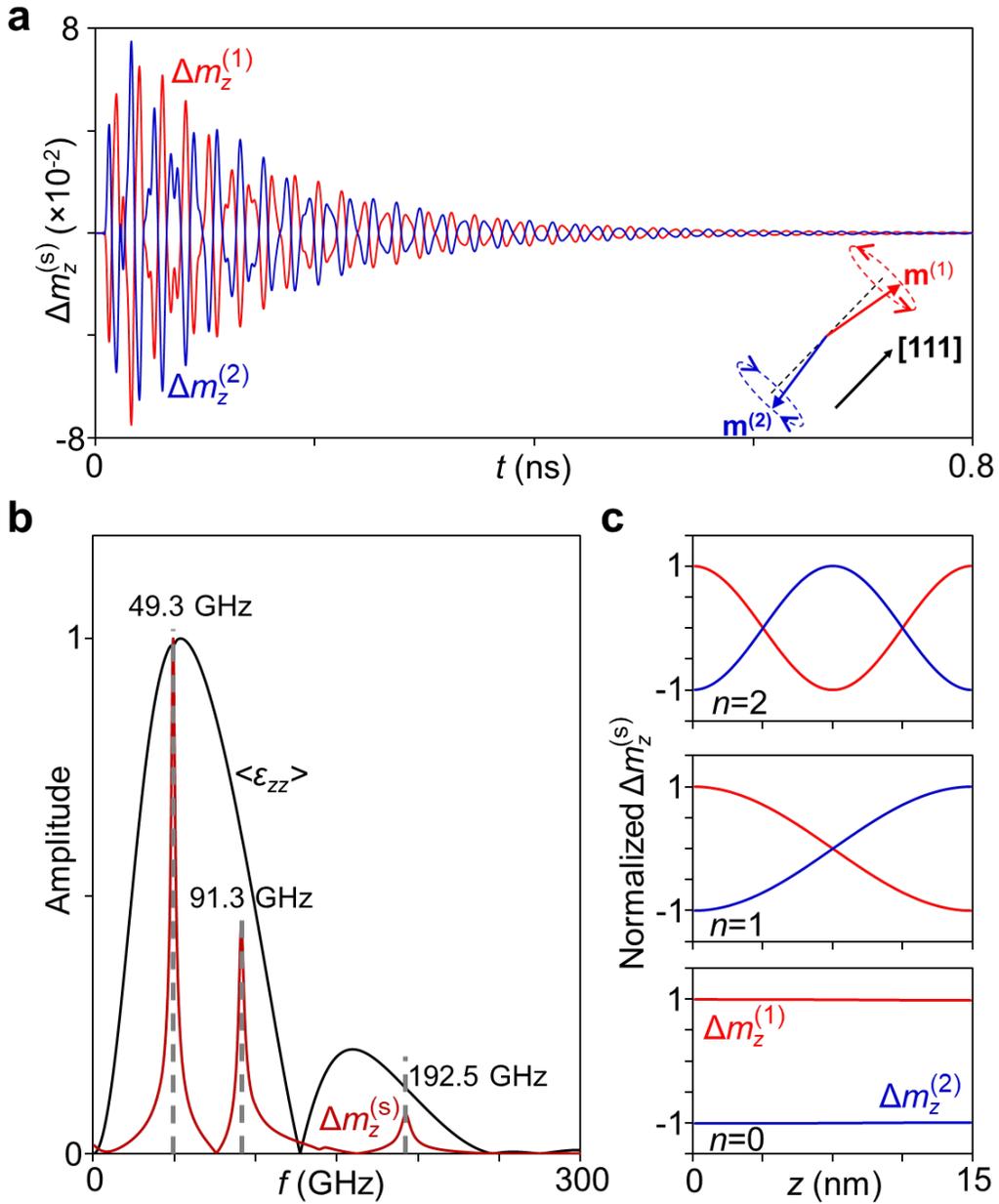

**Figure 5. AFM Magnon excitation.** (**a**) Temporal profiles of the $\Delta m_z^{(s)}(z=d,t)$ ($s=1,2$) in both magnetic sublattices of the FeMn at its top surface within $t=0$-$0.8$ ns. Note that $\Delta m_z^{(s)}(z,t) = m_z^{(s)}(z,t) - m_z^{(s)}(z,t=0)$, and $t = 0$ is defined as the moment the acoustic pulse enters the FeMn film from its bottom surface ($z=0$). (**b**) Frequency spectrum of the $\Delta m_z^{(s)}(z=d,t)$ within $t= 0$-$0.8$ ns (red line) and the thickness-averaged acoustic pulse $\langle\varepsilon_{zz}\rangle(t)$ (black line). (**c**) Spatial profiles of the $n = 2$, 1 and 0 mode AFM magnon components, represented by the normalized $\Delta m_z^{(s)}(z,t)$ (red line for $s=1$ and blue line for $s=2$) across the 15-nm-thick FeMn.

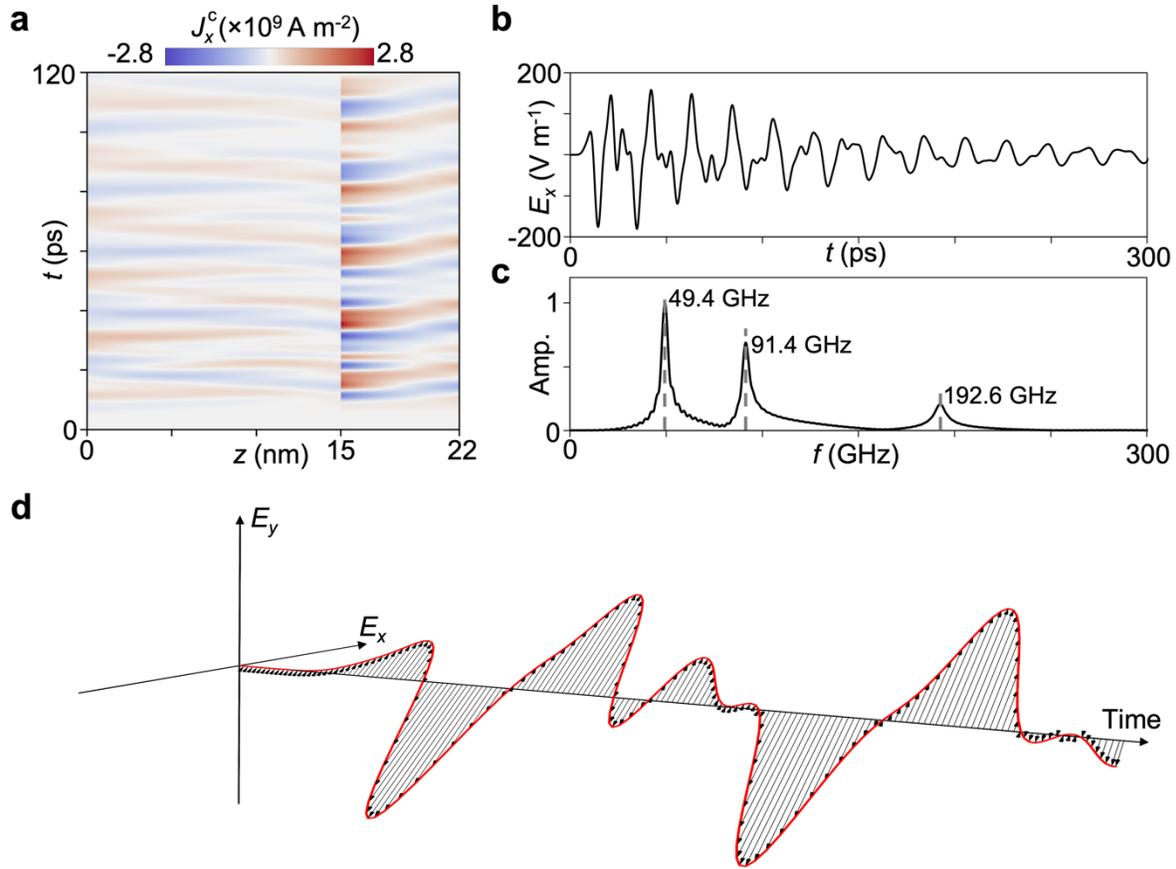

**Figure 6**. **AFM magnon detection.** (**a**) Spatiotemporal profile of the total charge current density $J_x^c$ in both 15-nm-thick FeMn and 7-nm-thick Pt. (**b**) Temporal profile of the emitted electric-field component $E_x(t)$ at 5 nm above the Pt top surface within $t = 0$-300 ps and (**c**) its frequency spectrum. (**d**) Evolution of the emitted electric field vector $\mathbf{E}(t)$ within $t$=0-50 ps, showing a linear polarization.